\documentclass[11pt,a4paper]{article}
\pdfoutput=1
\usepackage{a4wide}
\usepackage{jheppub}

\usepackage{epsfig}
\usepackage{latexsym}
\usepackage{amsfonts}
\usepackage{amsmath}
\usepackage{amsthm}
\usepackage{amssymb}
\usepackage{amsbsy}
\usepackage{multirow}
\usepackage{slashed}
\usepackage{color}
\usepackage{mathrsfs}
\usepackage{subfigure}
\usepackage[font={footnotesize,sl},bf]{caption}
\usepackage{wrapfig}

\newcommand{\bQ}{Q}

\def\calb         {{\cal B}}

\def\cale         {{\cal E}}
\def\calf         {{\cal F}}

\def\calj         {{\cal J}}

\def\call         {{\cal L}}

\def\caln         {{\cal N}}

\def\calt         {{\cal T}}

\newsavebox{\uuunit}
\sbox{\uuunit}
    {\setlength{\unitlength}{0.825em}
     \begin{picture}(0.6,0.7)
        \thinlines
        \put(0,0){\line(1,0){0.5}}
        \put(0.15,0){\line(0,1){0.7}}
        \put(0.35,0){\line(0,1){0.8}}
       \multiput(0.3,0.8)(-0.04,-0.02){12}{\rule{0.5pt}{0.5pt}}
     \end {picture}}

\newcommand{\beq}{\begin{eqnarray}}
\newcommand{\eeq}{\end{eqnarray}}

\newcommand{\be}{\begin{equation}}
\newcommand{\ee}{\end{equation}}
\newcommand{\bea}{\begin{eqnarray}}
\newcommand{\eea}{\end{eqnarray}}
\newcommand{\bean}{\begin{eqnarray*}}
\newcommand{\eean}{\end{eqnarray*}}

\def\x{\xi}

\newcommand{\ti}[1]{\tilde #1}

\def\to{\rightarrow}

\def\sF{{{ F}\!\!\!\!\hskip.8pt\hbox{\raise1pt\hbox{/}}\,}}
\def\som{{{ \omega}\!\!\!\!\hskip.8pt\hbox{\raise1pt\hbox{/}}\,}}
\def\sJ{{{\rm J}\!\!\!\!\hskip.8pt\hbox{\raise1pt\hbox{/}}\,}}

\newcommand{\bdm}{\begin{displaymath}}
\newcommand{\edm}{\end{displaymath}}
\newcommand{\nn}{\nonumber \\}

\newcommand{\QK}{{\cal Q}_K}

\begin{document}

\title{Structure of Six--Dimensional Microstate Geometries}
\author{Paul de Lange,}
\author{Daniel R.\ Mayerson,}
\author{Bert Vercnocke}
\affiliation{Institute for Theoretical Physics, University of Amsterdam,\\
Science Park 904, Postbus 94485, 1090 GL Amsterdam, The Netherlands}

\emailAdd{p.delange@uva.nl}
\emailAdd{d.r.mayerson@uva.nl}
\emailAdd{bert.vercnocke@uva.nl}

\abstract{We investigate the structure of smooth and horizonless microstate geometries in six dimensions, in the spirit of the five-dimensional analysis of Gibbons and Warner \href{http://arxiv.org/abs/1305.0957}{[arXiv:1305.0957]}. In six dimensions, which is the natural setting for horizonless geometries with the charges of the  D1-D5-P black hole, the natural black objects are strings and there are no Chern-Simons terms for the tensor gauge fields. However, we still find that the same reasoning applies:  in absence of horizons, there can be no smooth stationary solutions without non-trivial topology. We use topological arguments to  describe the Smarr formula in various examples: the uplift of the five-dimensional minimal supergravity microstates to six dimensions, the two-charge D1-D5 microstates, and the non-extremal JMaRT solution. We also discuss  D1-D5-P  superstrata and confirm that the Smarr formula gives the same  result as for the D1-D5  supertubes which are topologically equivalent.
}
\keywords{Black Holes in String Theory, Black Holes}

\maketitle

\tableofcontents

\section{Introduction}

The fuzzball programme argues that extended  objects of string theory alter the horizon of black holes quite drastically.\footnote{For reviews, see \cite{Bena:2007kg,Mathur:2008nj,Balasubramanian:2008da,Skenderis:2008qn,Chowdhury:2010ct,Bena:2013dka}. Related arguments instigated the recent firewall discussion  \cite{Braunstein:2009my,Mathur:2009hf,Almheiri:2012rt}.}  
Classical solutions to the supergravity equations of motion have played a significant role in shaping fuzzball discussions. These `microstate geometries' are smooth, globally hyperbolic, and solitonic stationary solutions that carry the same charges as a black hole and are argued to correspond to the gravitational interpretation of certain black hole microstates. 

The very existence of such smooth solitonic solutions was a bit puzzling. Based on Smarr formulas, many no-go theorems have been proven in the past for regular, stationary solutions in four dimensions, also in supergravity \cite{Breitenlohner:1987dg}. They can be summarized through the slogan: \emph{no solitons without horizons}. 
However, based on explicit construction, it has become clear that there are many supersymmetric horizonless smooth geometries with charges supported by topological fluxes. Also, families of near-supersymmetric microstate geometries are supported by fluxes on non-trivial two-cycles \cite{Bena:2011fc,Bena:2012zi}. 

The seemingly paradoxical existence of these horizonless microstate geometries was further clarified by Gibbons and Warner in \cite{Gibbons:2013tqa}. They revisited the Smarr formula in detail for five-dimensional asymptotically flat supergravity solutions and explained the mechanism that supports mass in a stationary, horizonless soliton. Besides receiving contributions from horizons, the Komar mass is also affected by previously neglected terms arising from spatial sections with non-trivial second cohomology. This is possible due to the existence of Chern-Simons terms in the action and cohomologically supported fluxes. The key slogan must be modified to \emph{no solitons without horizons or topology and fluxes}, which then holds regardless of supersymmetry and is in particular also valid for stationary solutions carrying the charges of a black hole with non-zero Hawking temperature. This has been further corroborated by the similar M-theory analysis of \cite{Haas:2014spa} and its compactification to five dimensions.

In this paper we want to discuss the topological Smarr formula in the six-dimensional arena appropriate to current microstate geometry research for the three-charge black hole. In addition, we want to explore non-extremal solutions, which have not been discussed before from the viewpoint of considering their topological contributions to the mass.\footnote{With the exception of the BPS-bound violating solution of \cite{Compere:2009iy} discussed in \cite{Gibbons:2013tqa}.}

The three charge black hole has five non-compact dimensions. It can, for instance, be obtained in IIB string theory on $T^4\times S^1$ with D1-D5-P charges on the internal directions. The string theory microstates dual to this black are given in the D1-D5 orbifold  CFT.  The discussion of the Smarr formula of \cite{Gibbons:2013tqa} concentrated on the five-dimensional microstate  geometries, which fit in five-dimensional $\mathcal{N}=1$ supergravity with three vector fields in $U(1)^3$, after dimensional reduction on $T^4\times S^1$.  However, the geometric interpretation of the D1-D5-P microstates can in principle excite the full ten-dimensional spacetime and does not have to allow a five-dimensional truncation.
Indeed, the microstate geometries for the two-charge D1-D5 black hole, or `supertubes',  depend on functions of one variable, the coordinate along the $S^1$. These functions describe a profile in the 8 spatial components orthogonal to $S^1$ (four non-compact dimensions \cite{Lunin:2001fv,Lunin:2001jy,Lunin:2002iz} and four torus directions \cite{Kanitscheider:2007wq}) and all IIB supergravity fields are excited.

A similar story is expected to hold for the microstates of the three-charge D1-D5-P black hole.  As argued in \cite{Bena:2011uw}, the generic microstate geometries in the  D1-D5-P frame are expected to be described by so-called \emph{superstrata}.  These should be obtained from adding momentum modes on two-charge D1-D5 supertubes and depend on functions of two variables. Their solution space might even give a leading contribution to the Bekenstein-Hawking entropy of the black hole \cite{Bena:2014qxa}.
Arbitrary superstrata excitations are most likely computationally beyond our reach and therefore the search for solutions has been focused on keeping the $T^4$ rigid. 
The rigourous treatment of \cite{Giusto:2013rxa} shows that any solution sharing the supersymmetries of the D1-D5-P brane system in IIB then fits in six-dimensional $\caln=(1,0)$ supergravity with  \emph{two} tensor multiplets, confirming earlier results of \cite{Kanitscheider:2006zf,Kanitscheider:2007wq,Giusto:2011fy,Giusto:2012jx}\footnote{The earliest three-charge geometries were six-dimensional by construction \cite{Lunin:2004uu,Giusto:2004id,Giusto:2004ip},  but lacked the intricacy  of  superstrata that has the potential of a solution space  with large entropic contribution.}. This  may be somewhat surprising, since the D1-D5-P black hole and many of its microstate geometries only excite one tensor multiplet. Nonetheless, this extra tensor multiplet helps to smoothen singularities in supergravity solutions \cite{Bena:2013ora,Bena:2014rea} and was in fact a key ingredient for the first successful construction of a superstratum \cite{Bena:2015bea}.



%

We extend the five-dimensional results of \cite{Gibbons:2013tqa} to the framework relevant for the more abundant conjectured six-dimensional solutions. The Komar integral that gives the conserved charge for a Killing vector again has a contribution for non-trivial topology, depending on the third cohomology of space. There are several new features in our discussion:
\begin{itemize}
\item \emph{New asymptotics:} The natural black object in our discussion is the six-dimensional D1-D5-P black string, where the string wraps the compact $S^1$ (compactification along $S^1$ gives the three-charge black hole). Hence we do not consider asymptotically flat $\mathbb{R}^{5,1}$ spacetimes, but rather focus on $\mathbb{R}^{4,1}\times  S^1$ asymptotics. This also implies a different relation between the Komar integral and Smarr formula. For  an extended object such as the black string, we cannot just relate the appropriate  Komar  integrals to the  ADM mass, but rather to combinations of both the integrated energy density and tension of the string \cite{Deser:1988fc,Stelle:1998xg,Townsend:2001rg}.

\vspace{-.2cm}
\item \emph{For supersymmetric solutions:} In six dimensions these always have a \emph{null} Killing vector  \cite{Gutowski:2003rg}  but not necessarily a timelike one as in five dimensions. The Komar integral for a null Killing vector does not give the ADM mass, but rather a `null charge'. We discuss the interpretation of this charge and its relation to the mass. As examples, we discuss the uplift of the 5D microstate geometries of \cite{Gibbons:2013tqa}, D1-D5 two-charge geometries, and D1-D5-P superstrata.  We find that the Komar integral for the  null charge is independent of the momentum charge P. This is  natural  as   D1-D5-P superstrata and D1-D5 supertubes share the same topology: both describe  ``wiggles'' of a  topological three-sphere.

\vspace{-.2cm}
\item \emph{For non-supersymmetric solutions:} We explore the JMaRT solutions \cite{Jejjala:2005yu} which have a timelike Killing vector. The Komar integral gives the ADM mass. The JMaRT  solitons are only smooth in six dimensions (not in five or four). As in the D1-D5  solutions, the flux through  a non-contractible $S^3$ supports the charges.
\end{itemize}

The rest of this paper is organized as follows. In section \ref{s:Smarr} we discuss the Komar integrals and the Smarr formula. We revisit brane-like solutions with compact dimensions and their energy densities and tensions. We normalize the Komar integral in terms of these physical quantities and highlight the difference between using timelike and null Killing vectors. Then we discuss the Komar integral in six-dimensional supergravity. We illustrate the general formula with supersymmetric examples in section  \ref{sec:susy}, and the non-supersymmetric JMaRT solutions \cite{Jejjala:2005yu} in section \ref{sec:nonsusy}. The supersymmetric examples include the uplift of the five-dimensional multi-center solutions discussed in \cite{Gibbons:2013tqa} and the D1-D5 Lunin-Mathur  geometries  \cite{Lunin:2001fv,Lunin:2001jy,Lunin:2002iz}; we also comment on D1-D5-P superstrata. 
We end with a discussion in section \ref{sec:discussion}.
Appendix \ref{app:5d6d} contains the  details of the uplift to six dimensions of five-dimensional multi-center solutions and in Appendix \ref{app:T4red}  we give the details of the truncation of IIB supergravity to six-dimensional supergravity with two tensor multiplets.

\section{Smarr Formula in Six Dimensions}\label{s:Smarr}

We discuss Komar integrals, the relation to the energy and tension of a solution, a Smarr formula for smooth horizonless solutions using topology and their application to six-dimensional supergravity with tensor multiplets..

\subsection{Komar integrals}
Any Killing vector $K$ of a metric on a $D$-dimensional Lorentzian spacetime defines a conserved quantity through a Komar integral:
\be
\QK = \frac{1}{8\pi G_D}\int_{\partial V_{\infty}} \star d K = \frac{1}{8\pi G_D} \int_{\partial V_{\infty}}(\partial_\mu K_\nu - \partial_\mu K_\nu)  d\Sigma^{\mu\nu} \,,\label{eq:Komar1}
\ee
where we integrate over a closed spatial surface at infinity.  Killing vectors enjoy the property $\nabla^2 K_\mu = -R_{\mu \nu}K^\nu$. 
With the help of Stokes' theorem, we can then rewrite this as a bulk integral over a volume $V$ on a spatial hypersurface with boundary $\partial V_{\infty} \cup \partial V_{\rm int}$:
\be
\QK =  -\frac{1}{4\pi G_D} \int_V \star (K^\mu R_{\mu\nu} dx^\nu) -\frac{1}{8\pi G_D}\int_{ \partial  V_{\rm int}}dS^{\mu\nu}(\partial_\mu K_\nu - \partial_\mu K_\nu) \,.  \label{eq:Komar2}
\ee
For a spacetime with a timelike Killing vector $K$, one usually relates the Komar integral to the ADM mass. However, this is only valid for an energy-momentum tensor that asymptotically approaches that of a weak static dust source, with $T_{00} \gg T_{0i},T_{ij}$ and $\partial_0 g_{ij}=0$ asymptotically. For a string-like object spanning the $y$ direction, we expect that $T_{00}$ and $T_{yy}$ will be of the same order, so we need to slightly modify the story.

\subsection{ADM integrals}

We now review the relevant results of  \cite{Townsend:2001rg}. To relate the  Komar integral to physical quantities such as the ADM energy, we consider an energy-momentum tensor that has asymptotically $p+1$ dominating diagonal components $T_{00}, T_{aa}, a = 1\ldots p$ and $p< D-3$. We assume all other components of the energy-momentum tensor are subleading compared to these. We take the $p$  coordinates to be compact and consider the linearization around a flat metric, $g_{\mu\nu} = \eta_{\mu\nu} + h_{\mu\nu}$ with Minkowski  reference metric
\be
ds^2_D  = -dt^2  +  \sum_{a=1}^p dy^a dy^a  + \sum_{i = 1}^n  dx^i dx^i\,,\qquad n = D - p -1\,.
\ee
We write the Einstein equations as
\be
R_{\mu\nu} -\frac 12  R  g_{\mu\nu} = 8  \pi G_D T_{\mu\nu}\,.
\ee
The energy density $\cal E$, average tension $\cal T$ and angular momentum density $\cal J$ are 
\begin{align}
{\cale} =\,&\int d^n x \langle T_{00} \rangle\,,\\
{\calt} =\,&-\frac 1  p \sum_{a=1}^p\int d^n x \langle T_{aa} \rangle\,,\\
{\calj}_{ij} =\,&\int d^n x(x_i  \langle T_{j0}\rangle  - x_j \langle T_{i0}\rangle)\,.
\end{align}
with the average  over the compact space $\langle X \rangle =  1/  {V_p}\int d^p y X$.
From the linearized Einstein equations,  one can then deduce the relations to the linearized metric components  $h_{\mu\nu}$ \cite{Townsend:2001rg}:
\begin{align}
{\cale} =\,&- \frac  1 {16 \pi G_D (n-2)}\int_{\partial V_{\infty}} dS_i \partial_i(  (n-1)h_{00}  -  h_{aa})\,,\\
{\calt} =\,& - \frac 1 p\frac  1 {16 \pi G_D (n-2)}\int_{\partial V_{\infty}} dS_i\partial_i(  p\,h_{00}  - (n+p-2) h_{aa}) \,.
\end{align}
These are the formulae that relate the asymptotic expansion of an extended object (where $T_{aa}$ is not negligible compared to $T_{00}$) to its mass and tension. After dimensional reduction over the $p$ internal directions, the ADM mass in $D-p$ dimensions is given by $\cale$. The angular  momentum  density  can still  be read  off from the  off-diagonal metric components:
\be
g_{0i} =   \frac{16\pi G_D} { \Omega_{D-2}}   \frac{x^j J^{ji} }{\rho^{n}}~+~ \dots  \label{asympg3}\,,
\ee
where  $\Omega_{D-2}$ is the volume of the unit  $(D-2)$-sphere and $\rho$ the radius in  the four spatial dimensions.

\subsection{Normalization of the  Komar integrals}

We now discuss the relation of the Komar integral to the energy density and tension.

\paragraph{Timelike Killing vector.}

One readily shows that for a  timelike  Killing vector $K$ that asymptotes to $K_\infty=\partial_t$ , we  have the  normalization
\be
\cale - \frac p{(D-3)} \calt = -\frac 1 {16\pi G_D}\frac{(D-2)}{(D-3)} \int_{\partial V_{\infty}} dS_{\mu\nu} (\partial^\mu  K^\nu - \partial^\nu K^\mu)\,.
\ee
For $p=0$, we retrieve the usual relations between the  ADM mass $M = \cale$ and the asymptotic form of the metric components \cite{Peet:2000hn,Gibbons:2013tqa}
\begin{eqnarray}
g_{00} &=&  -1 +  \frac{16\pi G_D} {(D-2)\, \Omega_{D-2}}  \frac{M}{\rho^{D-3} } + \dots  \,,  \label{asympg1} \\
g_{ij}   &=&  \left(1  + \frac{16\pi G_D} {(D-2)\, (D-3)\, \Omega_{D-2}}   \frac  M{\rho^{D-3}} \right)  \delta_{ij}+ \dots  \,.\label{asympg2}
\end{eqnarray}

\paragraph{Null Killing vector.}

Most of this paper is concerned with supersymmetric solutions in six dimensions. For these, it is useful to discuss $p=1$ and  consider null coordinates:
\be
u = \frac{t-y}{\sqrt{2}}\,,\qquad v = \frac{t+y}{\sqrt 2}\,.\label{eq:ty_uv}
\ee
For a null  Killing vector $K$ that asymptotically becomes $K_\infty=\partial_u$, one finds: 
\begin{align}
 {\cale}+\calt  =\,& - \frac 1 {8\pi G_D}\frac{(n+p-1)}{(n-2)}\int_{\partial V_{\infty}} dS_{\mu\nu} (\partial^\mu  K^\nu - \partial^\nu K^\mu)\,.
\end{align}
Note that these results, as in \cite{Townsend:2001rg}, are in principle only valid for time-independent metric perturbations. Metrics with a null Killing vector $\partial_u$ do not in general have to be time-independent. However, the time-dependence of the metric is heavily constrained. Since we average (integrate) over the internal, compact direction $y$, the resulting averaged metric must be time-independent and the results for the Komar integrals remain valid.

The normalization of the Komar integral (\ref{eq:Komar1}), which we use in  a 6D supergravity context for strings ($p=1$), implies that:
\be \QK = -\frac12\left({\cale}+\calt\right).\label{eq:Komar_ET}
\ee

\subsection{Six-dimensional supergravity}

Here we discuss the six-dimensional setup  relevant for the three-charge black hole. First we consider an arbitrary number $n_T$ of tensor multiplets; for superstrata in six  dimensions, $n_T =2$. We also explicitly give the formulas for $n_T=1$, which is relevant for all of the examples we discuss except the superstrata of section \ref{sec:superstrata}.

\subsubsection{Minimal supergravity with $n_T$ tensor multiplets}\label{sec:moretensors}

The six-dimensional supergravity theories of relevance to this work have an  $SO(n,m)$ global symmetry, with $n$ the number of tensors in the gravity multiplet. In the D1-D5-P frame, the relevant six-dimensional  theories are obtained by a compactification on $T^4$ or K3, which respectively give $\mathcal{N}=(2,2)$-supergravity with $SO(5,5)$ global symmetry and $\mathcal{N}=(2,0)$-supergravity with an $SO(5,21)$ symmetry group. 

Luckily, we do not need the full details of these extended supergravity theories. Rather, we can  consider  a consistent truncation to `minimal' six-dimensional supergravity with only $\mathcal{N}=(1,0)$ supersymmetry.  This theory has  $SO(1,n_T)$  global  symmetry where $n_T$ is the number of tensor multiplets and is in principle arbitrary as it is unfixed by supersymmetry. For our purposes, $n_T$ will be either 1 or 2, see  appendix  \ref{app:T4red} for more  details on the reduction from 10D.
Even though we focus on the theory  with $SO(1,n_T)$ global symmetry, our  results and in particular the Komar integrals \eqref{eq:thenullchargemoretensors} and \eqref{eq:thenullchargemoretensors2} below are straightforwardly extended to the bosonic sector of  six-dimensional supergravity theories with more supersymmetry, by formally  replacing the $SO(1,n_T)$ metric $\eta_{rs}$ with the metric of the appropriate global symmetry group.

When $n_T>1$, the equations of motion of the tensor fields do not follow from an action. We can still consider the `pseudo-action' \cite{Ferrara:1997gh,Riccioni:2001bg} for the bosonic fields\footnote{To avoid  confusion with standard notation $H$ for harmonic forms, we do not follow the notation of \cite{Ferrara:1997gh,Riccioni:2001bg} for the three-forms and the kinetic matrix.  To convert, use  $G^r  = H^r_{theirs}$ and  ${\cal M}_{rs} = (G_{rs})_{theirs}$.}
\be
\mathcal{L} = \frac14 R - \frac12  \partial_\mu v_r \partial^\mu v^s - \frac{1}{12} {\cal  M}_{rs}G^r_{\mu\nu\rho}G^{s\,\mu\nu\rho},\label{eq:6dsugra_more_tensors}
\ee
that captures the equations of motion of the scalar fields and the metric. The  scalars  parametrize the coset $SO(1,n_T)/SO(n_T)$. They   can be organized in the $SO(1,n_T)$-matrix $V = \begin{pmatrix}v_r\\x_r^M\end{pmatrix}$ with $M  = 1\ldots n_T$ and  $r=0\ldots n_T$.\footnote{It is customary to write the $SO(1,n_T)$ conditions $V\eta V^T= V^T  \eta V = \eta$ in component notation as $v_r v^r = 1, v^r  x^M_r=0, v_r v_s  - x_r^M  x_s^M = \eta_{rs}$.} They enter the tensor  dynamics through the scalar metric ${\cal M} =  \eta V^T V \eta$,  with $\eta$ the  $SO(1,n_T)$-metric, or in index notation 
\be
{\cal M}_{rs} = v_r  v_s  + x_r^M x_s^M\,.
\ee
The dynamics of the $n_T +1$ tensor fields $G^r$ are captured  by the self-duality conditions and Bianchi identities
\be
{\cal M}_{rs} G^s =  \eta_{rs} \star G^s\,,\qquad dG^r = 0\,,\label{eq:3form_duality}
\ee
where $\star$ is the six-dimensional Hodge star operator.
Finally, the Einstein equations are:
\be
R_{\mu\nu}=2\partial_\mu v^r\partial_\nu v_r+\frac 12 {\cal M}_{rs}G_{\mu\alpha\beta}^rG_{\nu}^{s\hspace{2pt}\alpha\beta}\,.\label{eq:EE_moretensors}
\ee



\subsubsection{Smarr Formula}

We are concerned with field configurations that respect the symmetry of a Killing vector $K$. This means the Lie derivative of the fields with respect to $K$ vanishes:
\be
{\cal  L}_K   g_{\mu\nu} = 0\,,\quad  
{\cal L}_K v^r = 0\,,\quad {\cal L}_K G^r = 0\,.
\ee
Since $dG^r$ and $\mathcal{L}_K = d\ i_K + i_K\ d$, we  can write the three-form and its dual as
\begin{align}
\label{eq:defH_more_tensors} i_K G^r  &= d \Lambda^r  + H^r\,,
\end{align}
for some globally defined one-forms $\Lambda^r$ and closed but not exact two-forms $H^r$. The Einstein equations \eqref{eq:EE_moretensors} become
\be 
K^{\mu}R_{\mu\nu} =    \frac12 \nabla_{\rho}\left({\cal  M}_{rs} \Lambda_{\sigma}^r G^s_{\nu}{}^{\rho\sigma} \right) + \frac12\left({\cal  M}_{rs}  H_{\rho\sigma} G^s_{\nu}{}^{\rho\sigma} \right)\,.
\ee
Then the Komar integral \eqref{eq:Komar2} is:
\be
\QK=
\label{eq:thenullchargemoretensors}
-\frac{1}{8\pi G_6}\int_V    {\cal  M}_{rs}H^r_{\rho\sigma}G_\nu^{s\hspace{2pt}\rho\sigma}dV^\nu
-\frac{1}{8\pi G_6}\int_{\partial V_{int}}\left(  {\cal  M}_{rs}\Lambda_\sigma^rG_{\mu\nu}^{s\hspace{5pt}\sigma}dS^{\mu\nu} +(\partial_\mu K_\nu - \partial_\mu K_\nu)\right).
\ee
As in \cite{Gibbons:2013tqa}, we find that   we can support  matter (non-zero Komar integrals) with horizons  or   with topology. For trivial topology,  $H ^r =0$ and the  Smarr formula \eqref{eq:thenullchargemoretensors} relates the Komar integral  to horizon quantities (area, charges and angular momenta). If also no horizons are present, the right-hand side  of \eqref{eq:thenullchargemoretensors} is zero and we get a vanishing Komar integral for the Killing vector  $K$. 

We are interested in spacetimes without inner boundaries. With \eqref{eq:3form_duality}, we  find
\begin{align}
\label{eq:thenullchargemoretensors2}\QK =& -\frac{1}{4\pi G_6} \int_V \eta_{rs} H^r \wedge  G^s \,,
\end{align}
so that only non-trivial topology can allow for non-zero Komar integrals.

\subsubsection{One tensor multiplet}

For many of the  solutions in this paper we can restrict to $SO(1,1)$ supergravity with $n_T=1$.
Including only one extra tensor multiplet in addition to the minimal supergravity multiplet is convenient as it allows for a Lagrangian description of the theory. The single self-dual three-form $G^+$ of the gravity multiplet can be combined with the single anti self-dual three-form $G^-$ of the tensor multiplet in one unconstrained three-form $G=\tfrac{1}{2}(G^+ + G^-)$.
The action becomes
\be
\mathcal{L} = \frac14 R - \frac12 \partial_\mu X  \partial^\mu X - \frac{1}{12} e^{2\sqrt{2}X} G_{\mu\nu\rho}G^{\mu\nu\rho}.\label{eq:6dsugra}
\ee
We introduce the dual three-form   (equivalent to  \eqref{eq:3form_duality}):
\be 
\tilde{G} = e^{2\sqrt{2}X} \star  G\,.\label{eq:dualG}
\ee
To  compare to the discussion of section \ref{sec:moretensors}, we can choose  $G^0  = G, G^1 = \tilde G$. The $SO(1,1)$  metric is then $\eta=\sigma_1$, and one  can choose the $SO(1,1)$ scalar matrix as $V = \exp (\sqrt2 X\sigma_3)$, where $\sigma_i$ are the  Pauli  matrices.

The Einstein equation  can be (re)written as:
\begin{align}
R_{\mu\nu} &= 2\partial_{\mu}X\partial_{\nu}X + \frac12\left(e^{2\sqrt{2}X} G_{\mu ab}G_{\nu}^{\ ab} + e^{-2\sqrt{2}X} \tilde{G}_{\mu ab}\tilde{G}_{\nu}^{\ ab}\right)\,.\label{eq:Einstein1}
\end{align}
The Komar integral \eqref{eq:thenullchargemoretensors2} is then
\begin{align}
\label{eq:thenullcharge}\QK =& -\frac{1}{4\pi G_6} \int_V  \left(H \wedge  \tilde G + \ti  H  \wedge G\right)\,,
\end{align}
with the harmonic forms $H,\tilde H$ defined through
\be
i_K  G = d \Lambda + H\,,\qquad \tilde i_K \tilde G = d \tilde \Lambda + \tilde H\,\label{eq:defH}
\ee
for some global one-forms $\Lambda$.

\subsubsection{Supersymmetry}

Let us also mention the fermionic content of the $SO(1,1)$ theory. The gravity multiplet consists of $(e_\mu,\psi_\mu^\alpha,B_{\mu\nu}^+)$ with $B^+$ a self-dual tensor such that $G^+\equiv dB^+=\star G^+$.  The tensor multiplet consists of $(B_{\mu\nu}^-,\chi^\alpha,X)$ with $G^-\equiv dB^-=-\star G^-$. The supersymmetry transformations of the fermions are:
\begin{align}
\delta\psi_\mu^\alpha&=(\partial_\mu-\frac{1}{4}e^{\sqrt{2}X}G^+_{\mu\nu\sigma}\gamma^{\nu\sigma})\varepsilon^\alpha\, ,\\
\delta\chi^\alpha&=\frac{1}{2i}(\sqrt{2}\gamma^\mu\partial_\mu X+\frac{1}{6}e^{\sqrt{2}X}G_{\mu\nu\rho}^-\gamma^{\mu\nu\rho})\varepsilon^\alpha.
\end{align}
Given a Killing spinor $\epsilon^\alpha$ we can construct the bilinear vector:
\begin{align}
K_\mu\varepsilon^{\alpha\beta}=\bar{\epsilon}^\alpha\gamma_\mu\varepsilon^\beta,
\end{align}
which is always a \emph{null} Killing vector, $K\cdot K=0$. The supersymmetry equations imply (using the form notation $K\equiv K^{\mu}g_{\mu\nu}dx^{\nu}$):
\begin{align}
dK&= 2 e^{\sqrt{2}X} i_K G^+ = i_K(e^{\sqrt{2}X} G + e^{-\sqrt{2}X}\tilde{G})   \,,\\
i_KdX&=0\,,
\end{align}
since the self-dual part of $G$ is given by $G^+ = 1/2(G + e^{-2\sqrt{2}X}\tilde{G})$.
Using $i_K \star  G = \star(G \wedge K)$, this allows us to write the null charge associated with $K$ as
\begin{align}
\QK&=\frac{1}{8\pi G_6}\int_{\partial V_{\infty}}\star dK
=-\frac{1}{8\pi G_6}\int_{\partial V_{\infty}} \left(\tilde{G}+G \right)\wedge K\ ,
\end{align}
where we have assumed that $X=0$ at infinity, which we can always do for asymptotically flat spacetimes.
In the microstate geometries of section \ref{sec:susy}, we find that $\partial V_{\infty}=S^1\times S^3$, and the Killing vector $K$ projected on this spacelike surface is (proportional to) the isometry along the compact $S^1$. In the notation of the metric (\ref{eq:6DMETRIC}) below, $K=-dv$ at spatial infinity. This means we simply get:

\begin{align}
\QK=-\frac{L_v}{8\pi G_6}\int_{S^3}\left(\tilde{G}+G\right)=-\frac{L_v\pi}{4G_6}(Q_e+Q_m),
\end{align}
where $L_v=2\pi R_v$ is the size of the $S^1$ direction parametrized by $v$ (at constant time). This relation is thus simply the BPS condition in 6D relating the null charge associated to $K$ to the electric and magnetic charges of the solution.

\section{Supersymmetric Examples}\label{sec:susy}

We now analyze  in detail the null Komar integral for known smooth supersymmetric solutions to six-dimensional supergravity. The structure of supersymmetric solutions in 6D minimal supergravity was studied in \cite{Gutowski:2003rg} and including an additional vector multiplet  and one tensor multiplet in \cite{Cariglia:2004kk}. Using the Killing spinors of such supersymmetric solutions, one can always construct a null Killing vector which locally is $V=\partial_u$. The metric can then be shown to take the form:
\be  \label{eq:6DMETRIC}
ds_6^2 = -2H^{-1}(dv+\beta_i dx^i)[du+\omega_idx^i+\frac{\mathcal{F}}{2}(dv+\beta_i dx^i)] + Hdx_4^2,
\ee
where $dx_4$ is the line element on the 4D ``base space'' $\mathcal{B}$, the one-forms $\beta=\beta_i dx^i,\omega=\omega_idx^i$ only have legs on $\mathcal{B}$ and the functions $H,\beta_i,\omega_i,\mathcal{F}$ are in general functions of $v$ and all of the 4D base coordinates $x^i$. The conditions that these functions (and the three-form and scalar) must satisfy for supersymmetric solutions can be found in \cite{Cariglia:2004kk},  or \cite{Bena:2011dd} whose conventions and notation we follow.  Note  that the ansatz \eqref{eq:6DMETRIC} only holds for sections \ref{sec:expectations}-\ref{ssec:D1D5}, in section \ref{sec:superstrata} we extend the ansatz for two tensor  multiplets.

\subsection{General expectations}\label{sec:expectations}

It is instructive to first work out the ADM integrals $\cale$ and $\calt$ for the three-charge solutions of our interest.
Asymptotically,  the metric \eqref{eq:6DMETRIC} approaches that of the three-charge black string  for which $H = (Z_2  Z_3)^{-1/2}, \calf= -Z_1,\omega=0,\beta=0$ and $Z_i = 1+ Q_i/\rho^2$, with $\rho$ the standard radial coordinate of the 4D base $\calb  = \mathbb{R}^4$. 
The asymptotic metric perturbation  in the coordinates $t,y$ \eqref{eq:ty_uv} is
\be
h_{00} = \frac 12 \frac{Q_2 + Q_3  +  Q_1} {\rho^2} + O(\rho^3)\,, h_{yy} = \frac 12 \frac{-Q_2 - Q_3  +  Q_1} {\rho^2}+ O(\rho^3)\,.
\ee
and we find that
\be
{\cale} =\,\frac  {\pi L_y } {4 \pi G_6}\left(Q_2+Q_3+\frac12Q_1\right)\,,\qquad
{\calt} =\,\frac  {\pi L_y } {4 \pi G_6}\left(Q_2+Q_3-\frac12Q_1\right) \,,
\ee
with $y \sim  y + L_y$. Note that $\cale$ is the ADM mass after dimensional reduction over the $y$-circle.\footnote{Note that the dimensional reduction in section \ref{sec:fivedmicro} and appendix \ref{app:secupliftsusy} is a reduction over the spacelike $v$-circle, which will give a different resulting 5D ADM mass in terms of $Q_1$, see eq.\ \eqref{eq:ADM_5D}.}
Using \eqref{eq:Komar_ET}, we  anticipate that the Komar  integral will be:
\be
\QK = -\frac12(\cale +\calt) = -\frac{\pi L_y}{4 G_6} (Q_2   + Q_3)\,,
\ee
and does not involve the momentum charge $Q_1$.

\subsection{The uplift of  five-dimensional microstate geometries}\label{sec:fivedmicro}

As a warm-up, we  consider the uplift of five-dimensional microstate geometries. Komar integrals and Smarr formulae for those geometries were discussed at length in \cite{Gibbons:2013tqa}, hence we do not go into much detail here.
The solutions are completely smooth multi-centered solutions of the 5D STU model with three gauge fields $A^I$ ($I=\{1,2,3\}$) and three scalars $X^I$, constrained by $X^1 X^2 X^3=1$. The 5D Lagrangian is given by (\ref{eq:app:usualSTU}). The 6D theory of minimal supergravity coupled to one tensor multiplet \eqref{eq:6dsugra} gives exactly this STU model when dimensionally reduced to 5D. See appendix \ref{app:5d6d} for more details.

The 5D solutions that we are interested in are given by the metric \cite{Gauntlett:2004qy,Elvang:2004ds,Bena:2004de}:
\be \label{eq:GWmetric} ds_5^2 = -Z^{-2} (dt + k)^2 + Z ds_4^2, \qquad Z = (Z_1Z_2Z_3)^{1/3}.\ee
where the 4D base space $\mathcal{B}$ is Gibbons-Hawking: it is a $U(1)$ fibration with coordinate $\psi$ over flat $\mathbb{R}^3$. The solutions are then determined by specifying the poles of eight functions $V, K^I, L_I, M$, which are harmonic functions on $\mathbb{R}^3$. For instance, we have $Z_I = L_I + C_{IJK}K^J K^K /2V$ with $C_{IJK}=|\epsilon_{IJK}|$. These eight harmonic functions must satisfy stringent conditions in order for the full 5D spacetime to be completely regular and asymptotically flat \cite{Bena:2007kg,Gibbons:2013tqa}. 

The gauge potentials in 5D are:
\be \label{eq:GWAI} A^I = -Z_I^{-1}(dt+k) + B^I,\ee
where $B^I$ is a magnetic potential (only well-defined locally).
The scalars are given by:
\be \label{eq:GWXI} X^I = \frac{Z}{Z_I}.\ee
For asymptotically flat 5D spacetimes, we have asymptotically:
\be Z_I \sim 1 + \frac{Q_I}{4r} =1 + \frac{Q_I}{\rho^2} ,\ee
with $r$ the usual radial coordinate on $\mathbb{R}^3$ and $\rho = 4r$ is the radial coordinate on the four-dimensional base. In microstate geometry literature, the charges $Q_I$ are normalized through the asymptotic expansion of the electric field in 5D as
$F_{0\rho} \sim 2 \frac{Q_I}{\rho^3}$ and not with factors involving the volume of the three sphere that are more common from Gaussian integrals. This means that we have:
\be -\frac{1}{16\pi G_5} \int_{\partial V_{\infty}} \star_5 F_I = \frac{\pi}{4 G_5} Q_I\,.\ee
For the six-dimensional metric, scalar and  tensor solutions see eqs.\ \eqref{eq:6duplift}.

\subsubsection{The topology of the base}
The poles of $V$ (`centers') indicate where the $\psi$-fibre degenerates in the 4D base space (although the complete 5D spacetime is always completely smooth). Since the $\psi$-fibre degenerates at each center, we can construct non-contractible compact two-cycles in the 4D space, which are also compact two-cycles in the full 5D geometry. These two-cycles are constructed by taking the $\psi$-fibration over an arbitrary path in $\mathbb{R}^3$ between two centers. This completely determines the 5D homology structure of simply connected solutions. For $N=2p+1$ centers, the global topology is  that of a $p$-fold connected sum of $(S^2\times S^2)$ with a point removed, for  $N=2p$ centers the topology is $(\mathbb{R}^2\times S^2)\#(S^2\times S^2)\#\ldots \#(S^2\times S^2)$.\footnote{We only discuss $V=\sum_i q_i/|x-x_i|$ with $|q_i|=1$, such that the centers are smooth points in the full space, and $\sum_i q_i=1$, such that the space is asymptotically flat.}

The five-dimensional ADM  mass of these solutions can be written as \cite{Gibbons:2013tqa}
\be M_{ADM, 5D} = -\frac{1}{32\pi G_5} C_{IJK} \alpha^I \int_{\Sigma_4} F^J\wedge F^K = \frac{\pi}{4G_5}\alpha^IQ_I = \frac{\pi}{4G_5} (Q_1 + Q_2 + Q_3),\label{eq:ADM_5D}\ee
where $\alpha^I=1$ for asymptotically flat solutions and $\Sigma_4$ is a spacelike surface of constant time. The integral of $F^J\wedge F^K$ is computed ``entirely with cohomology'', by calculating the flux of the $F^I$ over the non-trivial compact two-cycles of the geometry as well as the intersection number of these two-cycles.

\subsubsection{The topology of the uplift}

The six-dimensional uplift of (\ref{eq:GWmetric}) is a non-trivial fibration of the new coordinate $v$.
From the expression for the three-form:
\be 2\, G = (X^3)^{-2}\star_5 F^3 + F^2\wedge (dv + A^1),\ee
we can easily see that we have:
\be 2\, i_K G = d\left( \lambda_2(dv + A^1)\right) + d\left( Z_1^{-1}Z_2^{-1}(dt+k)\right) + F^1,\ee
where we have defined $\lambda_I=Z_I^{-1}-1$. The form given in the first term, $\lambda_2(dv + A^1)$, is well-defined.
The second term is $Z_1^{-1}Z_2^{-1}(dt+k)$ and is also a well-defined form (as discussed in \cite{Gibbons:2013tqa}). This implies the cohomology split:
\begin{align} 2\, \Lambda &= \lambda_2(dv + A^1) + Z_1^{-1}Z_2^{-1}(dt+k),\\
 2 H &= F^1.
 \end{align}
Similarly, we can find $\ti  \Lambda,\ti H$ by switching the roles of $Z_2$ and $Z_3$ in the above expressions. Note that also $2\ti H = F^1$.

The null charge is then:
\begin{align}
\QK 
&= -\frac{1}{4\pi G_6}\int_V\left(H\wedge \tilde{G} + \ti  H\wedge G \right)\\
&= -\frac{1}{16\pi G_6} \int_{V} \left( F^1\wedge (F^3\wedge dv) + F^1\wedge (F^2\wedge dv)\right)\\
&=  \frac{L_v}{16\pi G_6} \int_{\Sigma_4} \left( F^1\wedge F^3 + F^1\wedge F^2\right) \\
&= -\frac{L_v\pi}{4 G_6}(Q_2 + Q_3),
\end{align}
where we used the cohomological computation of the integral $F^I\wedge F^J$ in 5D over $\Sigma_4$ \cite{Gibbons:2013tqa}, and $V=S^1(v)\times  \Sigma_4$. We see that the null charge is simply the sum of electric and magnetic (string) charges. Note that in five dimensions, $Q_1$ is on the same footing as $Q_{2,3}$, but in six dimensions it is a momentum charge and does not appear in the null charge $\QK$.

The analysis above shows us that we clearly still have non-trivial compact two-cycles in six dimensions which are given by the trivial uplift of the two-cycles of the five-dimensional solution. These are the cycles supporting the cohomological flux $H,\tilde H\sim F^1$. The $S^1$-fibration of the coordinate $v$ over the compact two-cycles of the five-dimensional geometry also introduces new non-trivial three-cycles. Over these cycles, the cohomology elements $F^{2,3}\wedge dv$ have non-zero flux.

However, this is not quite the end of the story. In 6D, we must also have a non-trivial three-sphere at infinity. Indeed, the (electric string) charge in 6D is defined as:
\be Q_e = \frac{1}{2\pi^2}\int_{S^3(\infty)} e^{2\sqrt{2}X}\star G,\ee
where $S^3$ is the $S^3$ at infinity perpendicular to the string which is along $v$. Since the equation of motion for the three-form is simply $d(e^{2\sqrt{2}X}\star G)=0$, this $S^3$ at infinity must be non-contractible to be able to support non-zero flux for smooth solutions free of singularities. Note that this non-trivial three-cycle is absent in the original 5D geometry. This can be explained by the fact that this three-cycle must be homologically equivalent to an $S^1(v)$ fibration over a two-cycle in the 4D base (which we mentioned above). These new (compared to 5D) non-trivial three-cycles in constant time-slices of the six-dimensional geometry are an interesting feature of the $S^1(v)$ uplift.

\subsection{D1-D5 microstate geometries and supertubes}\label{ssec:D1D5}

We are now ready to discuss the topology and the Komar integral for more generic solutions of the D1-D5-P system. In this section, we first focus on the D1-D5 supertube solutions of Lunin and Mathur  \cite{Lunin:2001fv,Lunin:2001jy,Lunin:2002iz}. As we explain in section  \ref{sec:superstrata}, the result \eqref{eq:D1D5Komar} for the  Komar integral is the same for more generic D1-D5 supertubes and D1-D5-P superstrata, since those describe wiggles of the D1-D5 supertube and are topologically equivalent.

The D1-D5 Lunin-Mathur geometries are solutions to six-dimensional supergravity with only one tensor multiplet:
\begin{align}
 \label{eq:D1D5metric} ds^2 &= -\frac{2}{\sqrt{Z_1Z_2}} (dv+\beta)(du + \omega ) + \sqrt{Z_1 Z_2}ds_4^2,\\
 e^{2\sqrt{2}X} &= \frac{Z_1}{Z_2},\\
 2B &= -Z_1^{-1} (du+\omega)\wedge (dv+\beta) +\gamma_2.
\end{align}
Here $ds_4^2$ is the 4D flat metric with coordinates $x_i$ ($i=\{1,\ldots,4\}$) and $a_1,\gamma_2,
\beta,\omega$ are forms on the 4-manifold. The D1-D5 microstate is completely determined by profile functions $g_i(v),i=1\ldots4$ with $0\leq v\leq L$. Certain important functions are given by (for the complete list of fields, see for example \cite{Giusto:2013rxa}):
\begin{align}
 Z_2 &= 1 + \frac{Q_5}{L} \int_0^L \frac{1}{|x_i-g_i(v')|^2} dv', &  Z_1 &= 1 + \frac{Q_5}{L} \int_0^L \frac{|\dot{g}_i(v')|^2}{|x_i-g_i(v')|^2} dv',\\
 A &= -\frac{Q_5}{L} \int_0^L \frac{\dot{g}_j(v') dx^j}{|x_i-g_i(v')|^2}dv',& dB &= -\star_4 dA,\\
 \beta &= \frac{-A + B}{\sqrt{2}},& \omega &= \frac{-A-B}{\sqrt{2}},\\
 d\gamma_2 &= \star_4 dZ_2.
\end{align}

Perhaps the easiest explicit profile is the once-wound circle, given by (with $L=2\pi R_y$):
\be g_1(v) = a \cos(v/R_y), \qquad g_2(v) = a \sin(v/R_y), \qquad g_3(v)=g_4(v)=0.\ee
Then we can parametrize the (flat) 4D metric as:
\be ds_4^2 = \frac{f}{r^2+a^2} dr^2 + f d\theta^2 + (r^2+a^2)\sin^2\theta d\phi^2 + r^2\cos^2\theta d\psi^2,\ee
and the above functions become:
\begin{align}
Z_1 &= 1+\frac{Q_1}{f}, & Z_2 &= 1 + \frac{Q_5}{f},\\
A &= -a \sqrt{Q_1Q_5} \frac{\sin^2\theta}{f} d\phi,& B &= -a\sqrt{Q_1Q_5} \frac{\cos^2\theta}{f} d\psi,\\
f &= r^2 + a^2\cos^2\theta,
\end{align}
where $Q_1=a^2R_y^2/Q_5$, and the D1-D5 string at $x_i=F_i(v)$ is now at $r=0,\theta=\pi/2$ ($f=0$).

\subsubsection{Topology and homology}
The topology of the D1-D5 system with once-wound circular profile was discussed in \cite{Lunin:2002iz}. Any D1-D5 geometry with profile $g'_i(v)$ that can be continuously deformed into a circle will share the same topology of $\mathbb{R}^2\times S^3$. At infinity we have an $S^3(\theta,\phi,\psi)$ of the 4D base, which deforms continuously to the non-trivial $S^3(\theta,\tilde{\phi},\tilde{\psi})$ in the interior with $\tilde{\phi}=\phi+t/R, \tilde{\psi}=\psi+y/R$, while $S^1(y)$ (keeping $\tilde{\psi}$ fixed) shrinks to zero size in the interior.

Hence we clearly have exactly one non-trivial three-cycle given by the three-sphere at infinity, and one non-trivial (non-compact) two cycle, given by the volume element of the $\mathbb{R}^2$ factor. The three-cycle is again needed in this singularity-free geometry in order for the geometry to be able to support non-zero three-form flux. The intersection number between the two-cycle and the three-cycle is simply $+1$ (with suitable orientations of the cycles).

\subsubsection{Cohomology and null charge}
For a general D1-D5 geometry, we have:
\begin{align}
 2\, i_k G &= d(Z_1^{-1} (dv+\beta))\\
 &= \frac{1}{\sqrt{2}}d\left(Z_1^{-1}(dy+B)+Z_1^{-1}(dt-A)\right).
 \end{align}
Note that there is no obvious easy split to be made by defining $\lambda_1 = Z_1^{-1}-1$ and splitting off terms proportional to $\lambda_1$. This is because the fibres $A,B$ typically have singularities on the string profile and/or in the origin. So we can  leave the well-behaved  one-form $\Lambda$ implicit:
\begin{align}
2\, H &\equiv i_K  G  - d\Lambda  = d(Z_1^{-1} (dv+\beta)) -d\Lambda,
\end{align}
since the integrals we  will perform are independent of $\Lambda$ anyway.
In the explicit example of the once-wound circular profile, we can easily see that 
\be \frac{1}{L_v} \int_{\mathbb{R}^2} H = \frac{1}{L_v}\left(\frac{L_v}{2}\right) = \frac12,\ee
where we integrate the $\mathbb{R}^2$ cycle from the string profile (at $r=0,\theta=\pi/2$) to $r=\infty$, and we used that $Z_1^{-1}(f=0)=0$ and $Z_1^{-1}(r=\infty)=1$.

We see that $H$ is the cohomological dual of the non-trivial two-cycle in the geometry, as expected. The harmonic part of the three-form $G$ and its dual $\tilde{G}$ are both proportional to the volume form of the non-trivial three-cycle $S^3$:
\be \frac{1}{2\pi^2}\int_{S^3(\infty)} G = Q_5, \qquad \frac{1}{2\pi^2}\int_{S^3(\infty)} \tilde{G} = Q_1,\ee
as these parts precisely define the D1 and D5 charges of the geometry. Putting this together gives for the null charge:
\begin{align}
\QK 
&= -\frac{1}{4\pi G_6}\int_{\mathbb{R}^2\times S^3}\left(H\wedge \tilde{G} + \ti  H\wedge G \right)\\
&=-\frac{1}{4\pi G_6} \left(\int_{\mathbb{R}^2} H\right)\left(+1\right)\left(\int_{S^3}\tilde{G}\right) - \frac{1}{4\pi G_6} \left(\int_{\mathbb{R}^2} \ti  H\right)\left(+1\right)\left(\int_{S^3}G\right)\\
&= -\frac{L_v\pi}{4 G_6}\left( Q_1 + Q_5 \right),\label{eq:D1D5Komar}
\end{align}
where we used the intersection number to split the integral into separate integrals over the non-trivial cycles.

\subsection{D1-D5-P superstrata}  \label{sec:superstrata}

The most general three-charge microstate geometries that fall within six-dimensional supergravity arise from reduction on a rigid $T^4$ \cite{Giusto:2013rxa}. These  solutions excite all IIB supergravity fields in ten dimensions (metric, Ramond-Ramond fields $C_{(0)},C_{(2)},C_{(4)}$, as well as $B_{(2)}$ and the dilaton $\phi_1$). The solutions can be interpreted as solutions in minimal supergravity in six dimensions coupled to \emph{two} tensor multiplets, see appendix \ref{app:T4red}. 

These solutions require extending the results of section \ref{ssec:D1D5} in two ways: considering an extra tensor multiplet, and adding the momentum charge P. Only then can we cover both  generic D1-D5  geometries with a rigid $T^4$ \cite{Kanitscheider:2007wq} and  the  D1-D5-P superstrata \cite{Bena:2015bea}.  However, these more general solutions are  topologically equivalent to the D1-D5 supertubes \eqref{eq:D1D5metric}. We will show that the Komar integral is unchanged.

The general superstrata solutions as given in \cite{Giusto:2012jx,Bena:2015bea}, in six-dimensional language, fit  within the ansatz \cite{Giusto:2013rxa,Bena:2015bea}:
\begin{align}
 ds^2 &= \frac{\mathcal{P}}{Z_1Z_2}\left( -\frac{2}{\sqrt{\mathcal{P}}} (dv+\beta) \left[ du + \omega + \frac{\mathcal{F}}{2}(dv+\beta)\right] + \sqrt{\mathcal{P}}ds_4^2   \right),\label{eq:6dmetric_superstrata}\\
 e^{2\phi} &= \frac{Z_1^2}{\mathcal{P}},\\
 \chi &= \frac{Z_4}{Z_1},\\
 2B &= -\frac{Z_2}{\mathcal{P}} (du+\omega)\wedge (dv+\beta) + a_1\wedge (dv+\beta)+\gamma_2,\\
 B' &= -\frac{Z_4}{\mathcal{P}} (du+\omega)\wedge (dv+\beta) + a_4\wedge (dv+\beta)+\delta_2,\\
 \mathcal{P} &= Z_1 Z_2 - Z_4^2,
\end{align}
where, similar to the D1-D5 ansatz, $ds_4^2$ is the 4D flat metric and $\beta,\omega,a_1,a_4,\gamma_2,\delta_2$ are forms on this 4D base. We refer to \cite{Giusto:2013rxa,Bena:2015bea} for the full set of supersymmetry equations and equations  of motion and only quote those that we need:
\be
d\gamma_2 =  \star_4 dZ_2\,, \qquad  d\delta_2 = \star_4 d Z_4\,.
\ee

The tensor $B$ comes from the dimensional reduction of $C_{(2)}$ while $B'$ descends from $B_{(2)}$ in 10D; the scalar $\phi$ is simply the 10D dilaton while $\chi$ is the 10D axion $C_{(0)}$. For more information on the dimensional reduction from 10D to 6D and the realization of the $SO(1,2)$ symmetry, see appendix \ref{app:T4red}. This ansatz reduces to the D1-D5 ansatz \eqref{eq:D1D5metric} when $Z_4=a_4=\delta_2=0$; the tensor multiplet parametrized by the fields $B',\chi$ is set to zero, truncating the $SO(1,2)$ theory down to $SO(1,1)$.

The tensor multiplet scalars $\tau=\chi+i e^{-\phi}$ parametrize the coset $SO(1,2)/SO(2)$.
While $B$ and its field strength $G=dB$ are unconstrained, the tensor $B'$ satisfies a duality relation. Indeed, the field strength:
\be G' = dB' -2 \frac{\chi}{e^{-2\phi}+\chi^2} dB,\ee
is anti self-dual in six dimensions:
\be G' = -\star G'.\ee
Thus, we find the correct tensor field content for the $SO(1,2)$ theory of minimal supergravity with two tensor multiplets. 

The null charge is given by (see also appendix \ref{app:T4red}):
\be \label{eq:QKforstrata} \QK = -\frac{1}{4\pi G_6} \int_V  \left(H \wedge  \tilde G + \ti  H  \wedge G\right) +\frac{1}{8\pi G_6} \int_V  \left(H' \wedge  G' \right),\ee
where $H,\tilde{H}$ are defined as in \eqref{eq:defH}, similarly $H'$ is the  harmonic part of $i_K G'$, and the dual form $\tilde{G}$ is now defined by:
\be
 \tilde{G} = \frac{e^{2\phi}}{1+e^{2\phi}\chi^2} \star  G.
\ee
For the superstrata of \cite{Bena:2015bea}, the terms in (\ref{eq:QKforstrata}) involving $G,\tilde{G}$ can easily be seen to give the same contribution $\sim(Q_1+Q_5)$ as for the D1-D5 microstates above. The term involving $G'$ does not contribute. It is easiest to realize this by seeing that $dB'$ and $\chi dB$ fall off too fast at infinity to have a non-zero integral $\int_{S^3_\infty} G'$; in essence, this is because $Z_4$ falls off faster at infinity than $Z_1$ or $Z_2$ (which give the $Q_1, Q_5$ contributions to the null charge as in the D1-D5 case above).\footnote{To see this  fall-off explicitly we  quote the behaviour for the most general D1-D5 supertube invariant under $T^4$ rotations. This has five profile components $g_i, i=1\ldots 4$ and $g_5$, and the fields are \cite{Giusto:2013bda}:  
\be
\begin{array}{lll}
& Z_2 = 1 + \frac{Q_5}{L} \int_0^{L} \frac{1}{|x_i -g_i(v')|^2}\, dv'\,,\qquad
&Z_4 = - \frac{Q_5}{L} \int_0^{L} \frac{\dot{g}_5(v')}{|x_i -g_i(v')|^2} \, dv' \,,\\
& Z_1 = 1 + \frac{Q_5}{L} \int_0^{L} \frac{|\dot{g}_i(v')|^2+|\dot{g}_5(v')|^2}{|x_i -g_i(v')|^2} \, dv' \,\qquad &d\gamma_2 = *_4 d Z_2\,\qquad  d\delta_2 = *_4 d Z_4~,\\
& A = - \frac{Q_5}{L} \int_0^{L} \frac{\dot{g}_j(v')\,dx^j}{|x_i -g_i(v')|^2} \, dv' \,\qquad &dB = - *_4 dA~, \\
& \beta = \frac{-A+B}{\sqrt{2}}\,\quad&\omega = \frac{-A-B}{\sqrt{2}}\,\quad{\cal F}=0\,,\quad a_1=a_4=x_3=0~,
\end{array}
\ee
An explicit example is a  round profile in the  $\mathbb{R}^4$  base  and a non-zero $g_5$ component:
\be
g_1(v) = a \cos(v/R_y), \qquad g_2(v) = a \sin(v/R_y), \qquad g_3(v)=g_4(v)=0,\qquad  g_5(v)  = -\frac bk \sin   (v/R_y)\,.
\ee
The D1-D5  seed solution of  \cite{Bena:2015bea} starts from such a profile.
Then we have that
\be
Z_1 =   1 + \frac{Q_1}f  + c_1  \frac{\sin^{2k} \theta \cos(2k  \phi)}{(r^2  + a^2)^{k}f}\,,\qquad  Z_2 = 1  + \frac{Q_5}f\,,\qquad Z_4 = c_4 \frac{\sin^k \theta \cos(k\phi)}{\sqrt{r^2   + a^2} f}\,,
\ee
where $c_1 = \frac{Q_1 a^2  b^2}{2  a^2  +  b^2}$   and $c_4  = \sqrt{\frac{Q_1    Q_5}{a+2 + b^2/2}} b a^k$  are constants.
Clearly  $Z_4$ falls off too fast for the $H'\wedge  G'$-term to contribute to the  Komar integral.
For superstrata solutions, we refer to \cite{Giusto:2012jx,Bena:2015bea}.
} 
We conclude that:
\be \mathcal{Q}_K = -\frac{L_v\pi}{4 G_6}\left( Q_1 + Q_5 \right),\ee
just as for the D1-D5 supertube.  

That the null charge gives the same result for D1-D5-P superstrata as for the D1-D5 supertubes is not so surprising from a topological point of view. The important  thing to note is that a generic superstratum solution has the same topology as the D1-D5 round supertube. Superstrata describe fluctuations on top of a topologically non-trivial $S^3$ (shape modes depending on two variables), just as generic two-charge supertubes describe  one-dimensional shape modes on the $S^3$. This is the same $S^3$ present for the round supertube discussed in section \ref{ssec:D1D5}, and therefore supertubes and superstrata have a similar topological three-cycle.

\section{Non-Extremal Example}\label{sec:nonsusy}

We now discuss the JMaRT solutions of \cite{Jejjala:2005yu}, which have an interpretation as microstate geometries of the five-dimensional overspinning three-charge black hole. In the IIB frame, these are smooth solitons, with a natural interpretation in six-dimensional supergravity after dimensional reduction on the compact $T^4$.

\subsection{Metric and gauge fields}
The solitons are obtained  by demanding the  metric ansatz appropriate for describing the non-extremal three-charge black hole to be smooth. Usually, the five-dimensional physical charges are quoted, which in this case are the ADM mass $M_{ADM,5D}$, the electric charges $Q_1,Q_5,Q_p$, and the two angular momenta  $J_\psi,J_\phi$:\footnote{Standard conventions in the literature are to take $G_5=\pi/4$, which would render the prefactor $L_y\pi/(4 G_6)=1$. As in the rest of the paper, we instead choose to keep the explicit factors of $G_6$ in all of the relevant formulae. We also choose a normalization for the $Q_I$ that is the same as the rest of the paper, instead of the usual normalization which would include a factor of $L_y\pi/(4 G_6)$ in the definition of the $Q_I$ as well.}
\begin{align}
M_{ADM,5D} &= \frac{L_y \pi}{4 G_6} \frac m2 \sum_I \cosh 2 \delta_I\,, \qquad &J_\psi =-\frac{L_y \pi}{4 G_6}m( a_1 c_1 c_2 c_3 - a_2 s_1 s_2 s_3)\,,\\
\bQ_I &=  \frac m 2 \sinh 2 \delta_I\,, &J_\phi =-\frac{L_y \pi}{4 G_6}m( a_2 c_1 c_2 c_3 - a_1 s_1 s_2 s_3)\,, 
\end{align}
given in terms of  parameters $m,\delta_1,\delta_5,\delta_p,a_1,a_2$ and with the notation  $s_i = \sinh \delta_i, c_i  = \cosh  \delta_i$. The supersymmetric limit is $m,a_1,a_2 \to 0,\,\delta_i \to \infty$ while keeping $\bQ_I,m/\sqrt{a_i}$ fixed. We note that the 6D ADM mass (for the asymptotically $\mathbb{R}^{4,1}\times S^1$ spacetime) is actually:
\be \label{eq:JMART:MADM6D} M_{ADM,6D} = \frac{L_y \pi}{4 G_6}\frac m2\left( \cosh 2\delta_1 + \cosh 2\delta_5 + 2\cosh 2\delta_p\right),\ee
so the contribution due to the momentum charge (which is the charge from the graviphoton in reducing from 6D to 5D) is different.

We  choose to  write the metric and gauge fields in the notation of \cite{Chowdhury:2013pqa}. 
The metric, scalar and gauge field in 6D are (note that $B = -C_2/2$, with $C_2$  the RR  two-form  of \cite{Jejjala:2005yu}):
\begin{align}
ds_6^2  =\,& \frac{1}{H_p (H_1 H_5)^{1/2}}\left[-H_m\left(dt+k\right)^2 + H_p^2\left(  (dy + B_p^m + \frac{c_p}{s_p} k) + \frac {c_p}{s_p}(H_p^{-1}-1) (dt+k)
\right)^2\right]\nonumber\\
 &+ (H_1 H_5)^{1/2}ds_4^2, \\
 e^{2\sqrt{2}X} =\,& \frac{H_1}{H_5},\\
 -2 B =\,& \frac{c_1}{s_1} dt\wedge dy- \frac{c_1}{s_1} H_1^{-1} (dt+k)\wedge dy -B_1\wedge dz -\frac{c_1 c_p}{s_1 s_p} H_1^{-1}dt\wedge dk\nn
 &  - \frac{s_p}{c_p} dt\wedge B_1 - \frac{c_1}{s_1} H_1^{-1} dt\wedge B_3+ m s_5 c_5 \frac{r^2+a_2^2+m s_1^2}{f H_1}\cos^2\theta d\psi\wedge d\phi\,.
\end{align}
where the quantities used are defined by:
\begin{align}
ds_4^2 &= f \left( \frac{r^2}{g} dr^2 +  d\theta^2+\sin^2\theta d\phi^2+\cos^2\theta d\psi^2\right)\nonumber\\
&+ H_m^{-1} \left(a_1\cos^2\theta d\psi + a_2\sin^2\theta d\phi\right)^2 - \left(a_2\cos^2\theta d\psi + a_1\sin^2\theta d\phi\right)^2,\\
k  =\,& \frac{m}{f}\left[ -\frac{c_1 c_5 c_p}{H_m}\left(a_1\cos^2\theta d\psi + a_2\sin^2\theta d\phi\right) + s_1 s_5 s_p\left(a_2\cos^2\theta d\psi + a_1\sin^2\theta d\phi\right)\right],\nonumber\\
B^{(i)}  =\,& \frac{m}{f H_m} \frac{c_1 c_5 c_p}{s_I c_I} \left(a_1\cos^2\theta d\psi + a_2\sin^2\theta d\phi\right).
\end{align}
Everything is built from the following functions:
\begin{align}
H_i &= 1 + \frac{m s_i^2}{f},&  H_m &= 1 - \frac{m}{f},\\
f &= r^2 + a_1\sin^2\theta + a_2^2\cos^2\theta,& g &= (r^2+a_1^2)(r^2+a_2^2)-mr^2 = (r^2-r_+^2)(r^2-r_-^2),\nonumber
\end{align}

The three-form is simply $G = dB$. The dual potential, $\tilde{G} = d\tilde{B}$ is then given by:
\begin{align}
 \tilde B =\,& B\textrm{ with } s_1\leftrightarrow s_5; c_1\leftrightarrow c_5; H_1\leftrightarrow H_5.
\end{align}

\subsection{Constraints for smooth solutions}
Smooth JMaRT solutions are determined for fixed charges $ Q_1,Q_5,Q_p$, by two  integers $m,n$. One can extend these to  include  $\mathbb{Z}_k$ orbifolds  with $k$  an integer. They have the following relations between their parameters:
\begin{align}
r_+^2 =\, & - a_1 a_2 \frac{s_1 s_5 s_p}{c_1 c_5 c_p}\,,\\
M =\,& a_1^2 + a_2^2 - a_1 a_2 \left[\frac{c_1^2 c_5^2 c_p^2 + s_1^2 s_5^2 s_p^2}{s_1c_1 s_5c_5 s_pc_p}\right].
\end{align}
The constant $t$ slices have the topology of $\mathbb{R}^2 \times S^3/\mathbb{Z}_k$. The non-contractible $S^3$ is spanned at the origin $r = r_+$ by the coordinates $\theta,\tilde \psi, \tilde \phi$, with the identifications
\be
\tilde \psi = \psi - \frac {s_p c_p}{a_2 c_1 c_5 c_p - a_1 s_1 s_5 s_p}y\,,\qquad \tilde \phi = \phi-\frac {s_p c_p}{a_1 c_1 c_5 c_p - a_2 s_1 s_5 s_p}y\,,
\ee
The following quantization conditions ensure that the identification $y\to y+ 2\pi R$ is a closed  orbit:
\be
\frac {s_p c_p}{a_2 c_1 c_5 c_p - a_1 s_1 s_5 s_p}R = m\,,\qquad \frac {s_p c_p}{a_1 c_1 c_5 c_p - a_2 s_1 s_5 s_p}R = n\,,
\ee
for integers $m,n$. 

The $\mathbb{R}^2$ factor has a smooth origin at $r=r_+$, where the $t=constant$ part of the metric has the form (up to irrelevant constant prefactors)
\be
ds^2|_{dt =0} = d\rho^2 + \frac{\rho^2}{R^2} dy^2\,,
\ee
with the identification $y \sim y + 2\pi R k$ and the radius given by
\be
R =\frac{M s_1 c_1}{\sqrt{a_1 a_2}} \frac{\sqrt{s_1c_1 s_5c_5 s_pc_p}}{c_1^2 c_5^2 c_p^2 - s_1^2 s_5^2 s_p^2}\,.
\ee

\subsection{Komar integral}

We want to study the  Komar integral, which  reduces for this topology to
\begin{align}
\QK =& -\frac{1}{4\pi G_6}\int_V  \left(H \wedge  \tilde G + \ti  H  \wedge G\right)=-\frac{1}{4\pi G_6}\left(\int_{\mathbb{R}^2} H \int_{S^3}  \tilde G +\int_{\mathbb{R}^2} \ti H \int_{S^3}  G\right)\label{eq:KomarJMaRT1}
\end{align}
The  non-contractible $S^3$ is homologically equivalent  to the one at infinity appearing in Gauss' law. Hence  we  can perform the $S^3$ integral at spatial infinity:
\begin{align}
\frac{1}{4\pi G_6}\int_{S^3(\infty)} G =\,&  -\frac{1}{8\pi G_6}\lim_{r\rightarrow\infty} \int  d\left[m s_5 c_5 \frac{r^2+a_2^2+m s_1^2}{f H_1}\cos^2\theta d\psi\wedge d\phi\right]\\
=\,& \left.-\frac{\pi}{4G_6}\lim_{r\rightarrow\infty} m s_5 c_5 \frac{r^2+a_2^2+m s_1^2}{f H_1}\cos^2\theta \right|_{\theta=0}^{\theta=\pi/2} = \frac{\pi}{4G_6}Q_5
\end{align}
To obtain the  $H$-integral,  we  can in principle split the interior product of the Killing  vector with the three-form as
\be
i_K G = d \Lambda + H\,.
\ee
However, for our purposes we do not need to do this explicitly: the integral of $i_K G$ and of $H$ are identical, as the contribution  of $d\Lambda$ for  $\Lambda$  a well-defined one-form cancels anyway. 

To make  contact with the supersymmetric limit later, we consider the Killing vector
\be
K = \partial_t + \alpha \partial_y\,.
\ee
with $\alpha$ a  constant.
Then we find that locally
\be
 d\omega \equiv i_K G|_{t  = const.} \,,\quad   \omega  = \frac{c_1}{s_1} H_1^{-1} \left(dy  + (\frac{c_p}{s_p} - \alpha) k   + B^{(p)}\right) - \frac{c_1}{s_1}  dy + \left(\frac{s_p}{c_p} - \alpha\right) B^{(1)}\,.\label{eq:dB}
\ee
The one-form $\omega$ is zero at infinity  and well-behaved at any finite distance, but note that it is not globally well-defined. The integral $\int_{{\mathbb R}^2} i_K G_3$ only receives a contribution from the origin $r=r_+$. A short computation  shows that for constant $\tilde \psi, \tilde \phi$:
\be
B^{(i)}|_{r=r_+} =- \frac{s_p c_p}{s_i c_i} dy\,,\qquad  k|_{r=r_+} =0\,.
\ee
and hence the  first bracket in \eqref{eq:dB} does not contribute in the $\mathbb{R}^2$--integral. The other terms give:
\be
\int_{\mathbb{R}^2} H = \int_{\mathbb{R}^2} i_K G  = -L_y\omega_y|_{r=r_+}  = L_y\left(\frac {c_1} {s_1}  + \frac{s_p^2 -  \alpha  s_p c_p}{s_1   c_1}\right)=  L_y \frac{M_1 + M_p-\alpha Q_p}{Q_1}\,,
\ee
using the notation  
\be
M_i  =  \frac{m}{2} \cosh  (2\delta_i)\,,
\ee
which gives the  contribution to the 5D ADM mass in the $i$-channel (so that $M_{ADM,5D} = (L_y\pi)/(4G_6)\sum_i M_i$).

In the end, we find that \eqref{eq:KomarJMaRT1} becomes
\begin{align}
\QK 
=\, & -\frac{L_y\pi}{4G_6}\left(\frac{M_5 + M_p -\alpha Q_p}{Q_5}  Q_5  + \frac{M_1 + M_p -\alpha Q_p}{Q_1} Q_1\right)\\
=\, & - \frac{L_y\pi}{4G_6}\left(M_1  + M_5 + 2(M_p  -\alpha Q_p)\right)\,.
\end{align}

For  $\alpha = 0$,  we have $K = \partial_t$  and we retrieve the 6D ADM mass (\ref{eq:JMART:MADM6D}) for the Komar charge $\QK$.
Note that each term of the second line contributes to the  $M_p$-channel.
Also, in a sense, the non-extremality resides only in the  integral  over  $H$; the  integrals over  $S^3$  of $G_3,\tilde G_3$ contribute the charge. For $\alpha = 1$, so that $K = \partial_t  + \partial_y$, the Komar charge in the supersymmetric limit becomes the usual null charge $\QK = -(L_y\pi)/(4G_6)(Q_1 + Q_5)$.

\section{Discussion and Outlook}\label{sec:discussion}

Fluxes on non-trivial topology can support stationary configurations. This is a feature much used in microstate geometries and explained in detail in \cite{Gibbons:2013tqa} for five-dimensional smooth microstates. We have explored the six-dimensional guise of this mechanism for horizonless solutions. The three-form field strengths of six-dimensional supergravity and the 2-, 3-cohomology play a crucial role and give a non-trivial contribution to the Komar integral \eqref{eq:thenullchargemoretensors2} and thus to the conserved charges.

Many other avenues remain unexplored. One interesting direction is to explore the Smarr formula and the role of topology for non-flat asymptotics. As we have seen, compact directions give brane-like interpretations to the Komar integrals in terms of energy and tension densities.  It would be interesting to understand the extension to asymptotic Anti-de Sitter spaces. In string theory, spaces of the asymptotic form $AdS_p \times S^q$ are very  common. For these geometries, one must take care to regulate the Komar integral and perform a suitable background subtraction for the infinite $AdS$ background contribution and render the Komar integrals finite. However, it does not seem that this subtraction term would be expressible in terms of an interesting topological integral. 
In six dimensions, supersymmetric microstate solutions have $AdS_3 \times  S^3$ core regions, and  one can reinterpret our results for these geometries in their own right. In fact, for the D1-D5 solutions of section \ref{ssec:D1D5} it is clear that the relevant (non-trivial) three-cycle will be the $S^3$ and the two-cycle will be the ($t=const.$) non-compact spatial two-cycle of $AdS_3$. In other words, besides the subtlety of background subtraction, the situation for these $AdS_3$ geometries will be entirely analogous to the solutions considered here.
Perhaps more enlightening would be $AdS_5 \times S^5$ asymptotics, the arena of smooth  LLM geometries \cite{Lin:2004nb}. The topological contribution to the Smarr formula for 1/16 BPS solutions might also shed light on possible smooth geometries with the asymptotics of the  Gutowski-Reall black hole \cite{Gutowski:2004ez,Gutowski:2004yv}.

Perhaps a similar discussion of topology can give us insight into the cosmological horizon.  A Smarr formula has been discussed in the past \cite{Gibbons:1977mu,Abbott:1982jh}, but there has not been a discussion within supergravity models, nor with focus on topology.  We leave such investigations,  for instance for the de Sitter-Schwarzschild  black hole, to future work.

One of the original motivations of this work was to understand how to discriminate between supersymmetric and non-supersymmetric smooth solution with non-zero Hawking temperature. The best studied example of the latter are the JMaRT solutions, which are smooth in six dimensions and hence fit in our  current study.\footnote{It would be interesting to study the various known five-dimensional non-extremal constructions, such as those based on JMaRT  \cite{Bossard:2014ola,Katsimpouri:2014ara,Banerjee:2014hza} and Bolt-like \cite{Bena:2009qv,Bossard:2014yta} solutions.}
They have an ergoregion, which gives rise to an instability \cite{Cardoso:2007ws} that has been connected to Hawking radiation \cite{Chowdhury:2007jx,Chowdhury:2008uj}. One might expect that the appearance of an ergoregion in non-extremal microstate geometries is crucial for their decay and the connection to non-extremal black holes. Then one might also expect that the ergoregion plays a role in the universal characterization of microstate geometries through the Komar integral, as topology-supported solitons. However, the ergosurface is not topological and hence does not play a special role in the Komar integral. Hence the appearance of an ergoregion in the gravitational back-reaction of the probe constructions \cite{Bena:2011fc,Bena:2012zi} remains an open question.

\section*{Acknowledgments}

We would like to thank D.\ Anninos, I.\ Bah, I.\ Bena, J.\ de Boer, S.\ Giusto, B.\ Niehoff, D.\ Turton, A.\ Van Proeyen, E.\ Verlinde and N.\ Warner for enlightening discussions.
The research of B.V. is supported by the European Commission through the Marie Curie Intra-European fellowship 328652--QM--sing and  B.V. is deeply grateful for the enormous support of Evelien Dejonghe.
P.dL.  most gratefully acknowledges support of the ERC through the  Advanced Grant EMERGRAV. This work is part of the research programme of the Foundation for Fundamental Research on Matter (FOM), which is part of the Netherlands Organisation for Scientific Research (NWO).

\appendix{
\section{Uplift of Five-Dimensional Multi-Center Solutions}\label{app:5d6d}

\subsection{General reduction}\label{sec:appred}
Reducing 6D minimal supergravity plus a tensor multiplet gives the STU model in 5D. The 6D metric $\hat{g}_{ab}$ decomposes into the 5D metric $g_{ab}$, a graviphoton $A^{1}_a$, and a scalar $\phi_2$. The three-form gives two gauge fields: $\hat{G}_{abc} \sim (\star_5 F^2)_{abc}$ and $\hat{G}_{ab6}\sim F^3_{ab}$. Finally, our 6D scalar gives a scalar in 5D $\hat{X}=\phi_1$. We can then reparametrize the 5D scalars $\phi_1,\phi_2$ to get the usual three constrained scalars $X^I$ of the STU model.

We use hats to denote 6D quantities in this section; unhatted quantities, such as indices, are 5D. We start with the 6D Lagrangian:
\be \sqrt{-\hat{g}}\, \mathcal{L}_6 = \sqrt{-\hat{g}}\left[ \hat{R} -2 \partial_{\hat{\mu}} X\partial^{\hat{\mu}} X - \frac13 e^{2\sqrt{2}\hat{X}} \hat{G}_{\hat{\mu}\hat{\nu}\hat{\rho}}\hat{G}^{\hat{\mu}\hat{\nu}\hat{\rho}}\right]. \ee

We call the (spacelike) coordinate along which we reduce $y$. The reduction ansatz for the metric is:
\be d\hat{s}^2 = e^{ \phi_2/\sqrt{6}} ds_5^2 + e^{-3 \phi_2/\sqrt{6}} (dy + A^1_a dx^a)^2,\ee
with inverse:
\be (\partial \hat{s})^2 = e^{-\phi_2/\sqrt{6}} (\partial s_5)^2 - 2e^{-\phi_2/\sqrt{6}} A^{1\mu}\partial_{\mu}\partial_y + (e^{3\phi_2/\sqrt{6}}+e^{-\phi_2/\sqrt{6}}(A^1)^2)\partial_y^2.\ee
The Einstein-Hilbert Lagrangian then reduces to:
\be \frac{1}{G_6} \sqrt{-\hat{g}}\hat{R} = \frac{1}{G_5} \sqrt{-g}\left[ R - \frac12 (\partial\phi_2)^2 -\frac14 e^{-4\phi_2/\sqrt{6} \phi_2} (F^1)^2\right],\ee
where $G_6 = L_y G_5$. Note that $\sqrt{-\hat{g}} = e^{\phi_2/\sqrt{6}}\sqrt{-g_5}$. 

The kinetic term for the 6D scalar $\hat{X}$ gives the contribution:
\be \frac{1}{G_6}\sqrt{-\hat{g}}\left[-2 \partial_{\hat{\mu}} X\partial^{\hat{\mu}} X\right] = \frac{1}{G_5}\sqrt{-g}\left[ -2(\partial \phi_1)^2 \right].\ee

Finally, reducing the three-form can be done most easily using form notation. The reduction ansatz is:
\be 2\, \hat{G} = e^{-2\sqrt{2}\phi_1+2\phi_2/\sqrt{6}}\star_5 F^3 + F^2\wedge (dy + A^1),\ee
which also implies:
\be 2\, \hat{\star}\, \hat{G} = e^{2\phi_2/\sqrt{6}}\star_5 F^2 + e^{-2\sqrt{2}\phi_1}F^3\wedge (dy+A^1).\ee
Then the reduction of the kinetic term is:
\be
  2\, e^{2\sqrt{2}X} \hat{\star}\hat{G}\,\wedge\,  \hat{G} =     dy\,\wedge\, \left[ \frac12 e^{-2\sqrt{2}\phi_1+2\phi_2/\sqrt{6}} F_3\wedge \star_5 F_3 +  \frac12 e^{2\sqrt{2}\phi_1+2\phi_2/\sqrt{6}}\star_5 F_2\wedge F_2 + F^3\wedge F^2\wedge A^1\right].
\ee

Summarizing, the reduction gives us the 5D Lagrangian:
\begin{align} \sqrt{-g}\, \mathcal{L}_5 &= \sqrt{-g}\left[ R - \frac12 (\partial\phi_2)^2 -\frac14 e^{-4\phi_2/\sqrt{6} \phi_2} (F^1)^2  -2(\partial \phi_1)^2\right.\nn
 & \left. 
 - \frac14 e^{2\sqrt{2}\phi_1+2\phi_2/\sqrt{6}} (F^2)^2 - \frac14 e^{-2\sqrt{2}\phi_1+2\phi_2/\sqrt{6}} (F^3)^2 \right] -\frac14 \epsilon^{\mu\nu\rho\sigma\lambda}A^1_{\mu} F^2_{\nu\rho} F^3_{\sigma\lambda}.
 \end{align}
To bring this to the usual STU form, we can define:
\begin{align}
 X_1 &= e^{2\phi_2/\sqrt{6}},&
 X_2 &= e^{-\phi_2/\sqrt{6}-\sqrt{2}\phi_1},&
 X_3 &= e^{-\phi_2/\sqrt{6}+\sqrt{2}\phi_1},
\end{align}
so that $X^1X^2X^3=1$, and the Lagrangian can be written as:
\be \label{eq:app:usualSTU} \mathcal{L}_5 = R - \frac14 \frac{1}{(X^I)^2}(F^I)^2 - \frac12 \frac{(\partial X^I)^2}{(X^I)^2} -\frac14e^{-1} \epsilon^{\mu\nu\rho\sigma\lambda}A^1_{\mu} F^2_{\nu\rho} F^3_{\sigma\lambda},\ee
with sum over $I=\{1,2,3\}$ implied. This is the usual form of the STU Lagrangian. We can also write this as:
\be \mathcal{L}_5 = R - \frac12 Q_{IJ} F^I_{\mu\nu} F^{J\,\mu\nu} -  Q_{IJ}\partial_{\mu} X^I  \partial^{\mu} X^J - \frac1{24}e^{-1} C_{IJK} \epsilon^{\mu\nu\rho\sigma\lambda}A^I_{\mu} F^J_{\nu\rho} F^K_{\sigma\lambda},\ee
where we have $C_{IJK}=|\epsilon_{IJK}|$ and:
\begin{align}
 \frac16C_{IJK} X^I X^J X^K &=1,\\
 Q_{IJ} &:= \frac92 X_I X_J -\frac12 C_{IJK}X^K,\\
 X_I &:= \frac16 C_{IJK} X^J X^K.
\end{align}

\subsection{Uplifting SUSY solutions}\label{app:secupliftsusy}
The most general 6D supersymmetric metric can be written as \cite{Gutowski:2003rg,Cariglia:2004kk}:
\begin{align}
 ds_6^2 &= -2H^{-1}(dv+\beta)[du+\omega+\frac{\mathcal{F}}{2}(dv+\beta)] + Hdx_4^2,\\
  &= -H^{-1}\mathcal{F}[dv+\beta+\mathcal{F}^{-1}(du+\omega)]^2 + H^{-1}\mathcal{F}^{-1}(du+\omega)^2+H dx_4^2.
\end{align}
The rewriting of the metric in the second line shows us that we can reduce along $v$ as long as it is a spacelike coordinate, i.e. $\mathcal{F}<0$ everywhere. The reduction gives us:
\begin{align}
 ds_5^2 &= -H^{-4/3}\mathcal{F}^{-2/3} (du+\omega)^2 + H^{2/3}(-\mathcal{F}^{1/3}) dx_4^2,\nonumber\\
 e^{-3\phi_2/\sqrt{6}} &= H^{-1}(-\mathcal{F}),\nonumber\\
 A^1 &= \beta + \mathcal{F}^{-1}(du+\omega).
\end{align}
We see that the 6D null coordinate $u$ becomes a timelike coordinate in 5D \cite{Gutowski:2003rg}.

With the metric, gauge fields and scalars in 5D given by (\ref{eq:GWmetric})-(\ref{eq:GWXI}), we can then identify the appropriate 6D quantities in terms of the 5D ones as follows:
\begin{align}
 \mathcal{F} &= - Z_1,&
 \omega &= k,&
 \beta &= B^1,&
 H &= (Z_2Z_3)^{1/2}.
\end{align}
For reference, the full 6D fields are given by:
\begin{align}
 ds_6^2 &= -\frac{1}{Z_1(Z_2Z_3)^{1/2}} (du+k)^2 + (Z_2Z_3)^{1/2}ds_4^2 + \frac{ Z_1}{(Z_2Z_3)^{1/2}}(dv - Z_1^{-1}(du+k) + B^1)^2,\nonumber\\
 e^{\sqrt{2}X} &= e^{\sqrt{2}\phi_1} = X_1^{1/2}X_3 = \frac{Z_2^{1/2}}{Z_3^{1/2}}.\nonumber\\
  2G &= X_3^{-2} \star_5 F^3 + F^2\wedge (dv  +A^1).\label{eq:6duplift}
\end{align}

\section{Rigid \texorpdfstring{$T^4$}{} Reduction of IIB and \texorpdfstring{$SO(1,2)$}{} Truncation}\label{app:T4red}
The reduction of IIB supergravity to six-dimensional $\mathcal{N}=(1,0)$ supergravity with 2 tensor multiplets goes in two steps. In a first step, reduction of the bosonic sector on a rigid $T^4$ gives a theory with $SO(2,2)$ global symmetry \cite{Duff:1998cr}. Then the compatibility with D1-D5-P supersymmetries as in \cite{Giusto:2013rxa} leads to the bosonic sector of the $SO(1,2)$ invariant supergravity.

First, we reduce IIB supergravity on a $T^4$, keeping only the components of the fields with indices over the remaining six dimensions. This gives us two dilatons (from the 10D dilaton $\phi$ and the breathing mode of the $T^4$); two axions (from the 10D axion $C_{(0)}$ and from the only relevant component of $C_{(4)}$), along with the two reduced three-forms coming from the potentials $C_{(2)}$ and $B_{(2)}$. The reduction ansatz is \cite{Lavrinenko:1998hf,Duff:1998cr}:
\begin{align}
 ds_{10,str}^2 &= e^{\phi_1/2}\left( e^{\phi_2/2}ds_6^2 + e^{-\phi_2/2}ds_{T_4}^2\right),& C_{(0)} &= \chi_1, \nonumber\\
\phi &= \phi_1,& C_{(2)} &= C_{(2)},\nonumber\\
B_{(2)} &= B_{(2)}, & C_{(4)} &= -\chi_2\, \textrm{vol}(T_4) + \cdots, 
\end{align}
where $ds_{T_4}^2$ and $\textrm{vol}(T^4)$ are the flat metric and flat volume element on $T^4$. The $\cdots$ in $C_{(4)}$ are other terms that follow from the self-duality condition $F_{(5)}=\star F_{(5)}$. Note that we use the IIB supergravity conventions as in \cite{Bena:2015bea}. The resulting 6D Lagrangian is \cite{Duff:1998cr}:
\begin{align} \mathcal{L}_{6D,SO(2,2)} &= R - \frac12 (\partial\phi_1)^2 - \frac12 (\partial\phi_2)^2- \frac12 e^{2\phi_1} (\partial\chi_1)^2- \frac12 e^{2\phi_2} (\partial\chi_2)^2\nonumber\\
 & -\frac{1}{12} e^{-\phi_1-\phi_2} H_{(3)}^2 - e^{\phi_1-\phi_2} \frac{1}{12} F_{(3)}^2 + \chi_2 H_{(3)}\wedge dC_{(2)},
\end{align}
with $F_{(3)}\equiv dC_{(2)} - C_{(0)}H_{(3)}$. This reduction/truncation has an $SO(2,2)\cong SL(2)_1\times SL(2)_2$ symmetry where each $\tau_i=\chi_i+i e^{-\phi_i}$ parametrizes an $SL(2)/SO(2)$ coset.
The $SO(2,2)$ is not a symmetry of the tensor Lagrangian, but rather of the equations of motion and Bianchi identities. Those can be written as Bianchi identities of an $SO(2,2)$ vector of field strengths  $G^r$ with components
\be
\begin{pmatrix}
G^1\\
G^2
\end{pmatrix}
\equiv
\begin{pmatrix}
dB_{(2)}\\
dC_{(2)}
\end{pmatrix}
\,,\qquad
\begin{pmatrix}
G^3\\
G^4
\end{pmatrix}
\equiv
\begin{pmatrix}
\frac{d \call_{6D}}{dG^2}\\
-\frac{d \call_{6D}}{dG^1}
\end{pmatrix}
=
 -e^{\phi_2}(i\sigma_2)\cdot {\cal M}_{1} \cdot   \begin{pmatrix}
\star G^1\\
\star G^2
\end{pmatrix} +\chi_2 \begin{pmatrix}
G^1\\
G^2
\end{pmatrix}
\,.
\ee
Those tensors  obey  the duality relation  (compare \eqref{eq:3form_duality}):
\be
 {\cal M}_{rs} G^s = \eta_{rs}\star G^s\,,
\ee
with the off-diagonal $SO(2,2)$ metric $ \eta = (i \sigma_2)\otimes (i \sigma_2)$ and scalar matrix 
\be
{ \cal M} = {\cal M}_2(\tau_2) \otimes {\cal M}_1 (\tau_1),\quad \text{with}\quad  {\cal M}_i = V_i V_i^T,\quad  V_i = \begin{pmatrix}e^{-\frac 12 \phi_i} &\chi_i e^{\frac 12\phi_i}\\ 0& e^{\frac 12  \phi_i} \end{pmatrix}.
\ee

It is important to realize that this $SO(2,2)$ theory cannot be the bosonic part of any supergravity theory. 
One can perform a further truncation to obtain a theory that can be the bosonic part of $SO(1,2)\cong SL(2)$ supergravity by setting $\tau_2=f(\tau_1)$ with  $f$  an $SL(2)$-transformation. This identifies a `diagonal' $SL(2)$ subgroup in $SO(2,2)\cong SL(2)_1 \times SL(2)_2$.  The four tensors $G^r$ then decompose in a singlet and a triplet under this truncation. Consistency of the truncation requires that we put the singlet to zero.

We are interested in solutions with the supersymmetries of the D1-D5-P system \cite{Giusto:2013rxa}, giving the truncation:
\be 
\tau_2 = -\frac{1}{\tau_1}.
\ee
The $\tau_2$ equation of motion then requires that we put the singlet $G^1 + G^4$ to zero.
The remaining three field strengths are 
\be
\hat G^1=\frac 12 ( G^3-  G^2)\,,\quad \hat G^2  = \frac 12  (G^2  + G^3)\,,\quad \hat G^3 = \frac 12 (G^4 - G^1) \,,
\ee
Dropping the hats again, $G^r$ then obeys the self-duality relation with the $SO(1,2)$ matrix 
\be
V = \exp(\chi  E_+)\exp (\phi H/2)\,,\qquad E_+ = \begin{pmatrix}0&0&1\\0&0&1\\1&-1&0\end{pmatrix}\,,\qquad  H = \begin{pmatrix}0&2&0\\2&0&0\\0&0&0\end{pmatrix}
\ee
and the Komar integral \eqref{eq:thenullchargemoretensors2} applies.

To make the connection to the theory with one tensor  multiplet clear, we write the vanishing singlet as an anti self-duality constraint on a three-form $G'$:
\be G' = -\star G' ,\qquad G' = dB_{(2)} + \chi_2 dC_{(2)} = \frac{e^{-2\phi_1}H_{(3)} - \chi_1 F_{(3)}}{e^{-2\phi_1}+\chi_1^2}.
\ee
We can then take $G\equiv (1/2)dC_{(2)}$ to be the (unrestricted) tensor that is the combination of the other self-dual and anti self-dual tensors. In section \ref{sec:superstrata}, we take $\phi=\phi_1,\chi=\chi_1$ and $B=(1/2)C_{(2)}, B'=B_{(2)}$.  An obvious further truncation of this $SO(1,2)$ theory is to take $G'=0,\chi_1=0$ which leaves us with the $SO(1,1)$ sector used in large parts of this paper, after the identification $\phi_1=\sqrt{2} X$. 

For the $SO(1,2)$ theory with the unrestricted three-form $G$ and the anti self-dual three-form $G'$ as defined above, the generalization (\ref{eq:thenullchargemoretensors}) of (\ref{eq:thenullcharge}) for the null charge reduces to:
\be \QK = -\frac{1}{4\pi G_6} \int_V  \left(H \wedge  \tilde G + \ti  H  \wedge G\right) +\frac{1}{8\pi G_6} \int_V  \left(H' \wedge  G' \right)     \, ,\ee
where $H,\tilde{H}$ are defined as in (\ref{eq:defH}), keeping in mind the $SO(1,2)$-generalized definitions for the dual form:
\be
 \tilde{G} = \frac{e^{2\phi_1}}{1+e^{2\phi_1}\chi_1^2} \star  G,
\ee
The harmonic  form $H'$ is defined by the split:
\be i_K G' = d\Lambda' + H',\ee
where $\Lambda'$ is a globally defined one-form.

\bibliographystyle{JHEP}
\bibliography{ergoballs}

\end{document}